# Agent-based environment for knowledge integration

Anna Zygmunt , Jarosław Koźlak, Leszek Siwik

Department of Computer Science, AGH University of Science and Technology,
Al. Mickiewicza 30, 30-059 Kraków, Poland
{azygmunt, kozlak, siwik}@agh.edu.pl

**Abstract.** Representing knowledge with the use of ontolgy description languages offers several advantages arising from knowledge reusability, possibilities of carrying out reasoning processes and the use of existing concepts of knowledge integration. In this work we are going to present an environment for the integration of knowledge expressed in such a way. Guaranteeing knowledge integration is an important element during the development of the Semantic Web. Thanks to this, it is possible to obtain access to services which offer knowledge contained in various distributed databases associated with semantically described web portals. We will present the advantages of the multi-agent approach while solving this problem. Then, we will describe an example of its application in systems supporting company management knowledge in the process of constructing supply-chains.

## 1   Introduction

The accessibility of different knowledge bases and especially the Internet network in its current form at present provides a huge amount of different information resources and numerous application services. An important challenge here is to guarantee the most efficient and user-friendly solution of the possibilities provided by this kind of environment.

This may be offered by the development of a Semantic Web that embraces information existing in the WWW network and by integrating it with a software in order to realize various activities in response to the demands and requirements of the user, taking into consideration their preferences and cooperating with such different modules distributed in the Internet. An elaboration of this kind of infrastructure will offer a new quality of Web use and will provide users a greater amount of knowledge and services that are described in a correct way and useful to them. To achieve this goal, it is necessary to solve many different problems at the same time: a representation of a semantic of web resources, to guarantee reasoning mechanisms to supplement the lack of or only partial knowledge, the composing of web services, an analysis of the knowledge gathered and what is available in different thematic portals using techniques of artificial intelligence such as data mining or machine learning.

We will focus particularly on presenting an infrastructure needed for leading the process of integration of knowledge expressed with the use of knowledge description languages [7,8] like OWL[4] and RDF [3]. We will present different kinds of knowledge integration, described in the literature. Then we will present an environment for integrating ontologically expressed knowledge. Special attention will be given to the application of a multi-agent approach to solve this problem. We will also present our pilot environment based on the multi-agent platform JADE.

## 2   Study of Problems

In our work we are going to focus on three important problems regarding the knowledge expressed using ontologies: reasoning, integration and application of the multi-agent process to perform the integration of the knowledge stored in distributed knowledge bases. In the following sub-chapters we will outline in short the characteristics of research carried out in these domains.

### 2.1 Ontology Reasoning

The fundamental principle of the reasoning process in expressing the knowledge using ontologies is Description Logics (DL)  [9;1] which define two knowledge components: TBox and ABox. The TBox (Terminology Box) component includes domain definitions (i.e. declarations of concept properties). TBox determines the subsumption relations and is implementation-independent. The knowledge stored in TBox can be characterised as persistent (it does not evolve over time). The ABox (Assertional Knowledge) component contains case-specific knowledge: it assigns specific meanings to concepts derived from TBox. The knowledge stored in ABox is inherently transient and dependent on circumstances. The basic reasoning mechanism which exists in ABox involves checking whether a given unit is included in a selected concept. More complex mechanisms exist as well (although they all follow on from the basic one):

### 2.2 Ontology Integration

As a result of integration, there is the possibility to obtain access to higher amounts of information, it is also possible to obtain additional information which results from relations between concepts present in different knowledge sources.
It seems that an integration of ontologies may provide additional qualities to the applications being developed. Unfortunately at present, there are no ready-to-use, simple and in-use solutions that could automatically control this process. It is possible to distinguish [11] several different schemes of ontology integration. In the *single ontology* approach, all the ontologies are integrated into one global one and a unified access to a knowledge model takes place. In the *multiple ontology* approach, each of the information sources possesses their own local ontology, which makes it possible to use a separate dictionary and the ontologies may be developed independently.

A combination of these two approaches is a hybrid approach where its information source has its own local ontology derived from a global ontology to ensure easier adjustment. Carrying out the ontology integration the following operations are usually used [6]: mapping, alignment merging and integrating. During the process of ontology integration the following elements of the problem solution exists: an analysis of the lexical and structural similarities, in the aim of finding concepts most appropriate to one another, a use of existing tools and knowledge bases to find the relations between the concepts and domains to which they can be rated. It is worth mentioning here that semantic dictionaries like WordNet are specially designed high-level ontologies

*Upper level ontology* is an attempt to create a general description of concepts and relations among them which may be the same and possible to use for different knowledge domains. The main goal of the creation of such a knowledge base is to make it possible to access different ontologies through the upper ontology. One of the most popular definitions of the approach based on upper ontologies is SUMO (Suggested Upper Merged Ontology) [10]

**2.3 Overview of Environments for Ontology Integration**

To make possible the efficient use of the knowledge included in distributed knowledge sources, it is necessary to possess an appropriate infrastructure. In our opinion, the multi-agent [5] approach offers important advantages which support a knowledge integration process. The multi-agent approach is based on the assumption that systems have a distributed and decentralised structure consisting of autonomous rational entities called agents that cooperate with one another to realise their own tasks. This subsequently results in the realisation of the global goal of the system. Additionally, agents are equipped with features such as: perception of the environment, capacity to perform actions which modify the environment, use of interaction protocols which describe the possible conversation flow between them and using agent communication languages which describe the structure of exchanged messages, the construction of plans that have as their goal the realisation of assumed goals and governing the process of the machine learning so that it can realise its task in the dynamically changing environment as best as possible. These features may be very useful when we create an infrastructure for knowledge integration.

Several multi-agent environments were performed that have the goal of support in the process of agent development, deployment and multi-agent interactions. The set of specifications FIPA has the goal of elaborating requirements which agents have to fulfil to make the cooperation among them possible (http://www.fipa.org). Particularly, these specifications contain multi-agent communication languages and protocols, agent-transport protocols, agent management rules, agent architectures and their application. The most popular agent platform is JADE (Java Agent Development framework) [2] . During the realisation of the system other tools and software for use of ontologies were applied: ontology editor Protege, a library for accessing ontology models JENA and to create mappings between the ontology model and information stored in the database (D2RQ).

## 3. Environment for Ontology Integration

Our work concerns an overview of problems which we encountered during the development of an infrastructure that supports reasoning using a semantic knowledge representation distributed in the Internet while taking advantage of the possibilities of different actions on the basis of knowledge described in such a way. This knowledge was described with the use of ontological description languages (like *OWL*).

In the presented solutions, we will focus our attention on the application of different integration techniques: working on the local level (concerning particular concepts and their attributes) algorithms of lexical and structural comparison or checking of similarity between larger parts of a graph with the use of *Similarity Flooding* algorithm. We also applied additional approaches based on a thesaurus for looking for synonyms or on the use of high level ontology – *Upper Ontology* (such as SUMO) to adjust concepts from the ontology to a given set of concepts which identify important notions.

The infrastructure takes advantage of the agent platform JADE. The agent infrastructure for ontology integration was based on the assumption that a knowledge expressed using ontology languages is accessible in the form of decentralized knowledge sources (fig. 1). The system has a distributed architecture, each node possess an program instance and own ontology with a database. We can distinguish three main kinds of agents: Container Agent (represents an node), DistributedQuery Agent (represents the queries sends to the system by users) and Distributed QueryAgent (supervises a ontology integration process on its node).

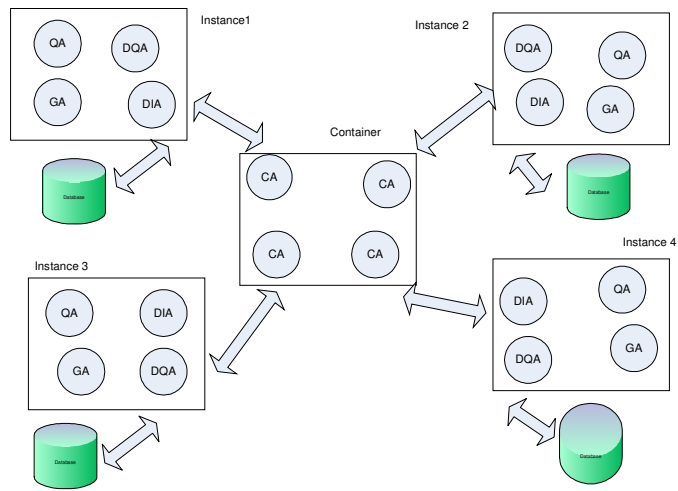

Fig. 1. General system architecture (CA – Container Agent, DIA – Distributed Integration Agents, QA – Query Agent, GA – Grade Agent)

Such architecture has many advantages: the possibility of information exchange between different centres, lack of necessity of possessing a knowledge about the instances of application currently accessible and the free flow of knowledge.

There are also disadvantages of this pilot realization. The costs of communication between the agents are high, during the integration phase the number of messages increases linearly with the increase of agent number.

**3.1 Agents in the System**

The agents are inheriting the functionality provided by *AbstractAg* class defined in JADE environment: There are following Agents in the *System*:

- *MasterAgent* – performs the necessary operations during program start, like the creation of other agents,
- *GradeAgent* – is an agent which performs integration using different methods, it also makes a choice of the most appropriate integration methods for a given problem and then sends a request to a special agent,
- *QueueAgent* –an agent which puts the integration requests into the queue,
- *ContainerAgent* – an agent which creates a distributed infrastructure, it makes the communications between the instances of the program possible,
- *DistributedIntegratingAgent* - after receipt of the query it sends its ontology to agent container,
- *DistributedQueryAgent* – this agent is created after the arrival of integration request, queries to ontology with the use of a functionality of JENA language, sends a requests to deliver ontologies for integration to all the needed system instances and then after receiving the ontologies, it sends them to *QueueAgent*,
- *IntegratingAgent* – integrates ontological model being sent to it, provides the main functionality for each integration agent, the special functionalities are provided by agents using a particular integration scheme,
- *MetricSimilarityIntegratingAgent* – compare instances using similarity metric,
- a set of different integrating agents, each of them is functioning according to its own algoriths, which use methods presented in section about integration methods: *PromptIntegrating Agent (*uses functionality offered by a Prompt tool), *SimilarityIntegratingAgen , JenaIntegratingAgent (*uses basic functions offered by Jena library*), DictionaryIntegratingAgent (*uses a dictionary of synonyms and looks for the suitable synonyms in the ontologies being integrated*)*, and the last three using extention integrating methods - *InstanceIntegratingAgent*, *InstanceSymmetricIntegratingAgent*, *InstanceJaccardIntegratingAgent.*

The integration of ontologies is coordinated by an agent called GradeAgent. This agent supervises a process of estimation of similarity of given classes/concepts with the use of different methods. It creates also another integrating agents, sends to them messages with models for evaluation, receives matrixes with results and constructs integration commands as well as initialises a final step of integration.

**3.2 Interactions Between Agents**

The realised environment is based on the cooperation of many agents that find things out from each other thanks to a repository. *Container Agents* are run on the server, their role is to represent original instances of the program. During the start of each instance two agents: *Container Agent* and *Queue Agent* (responsible for queuing requests for ontology integration coming from other instances) are created. At the moment of a request arriving, the following operations are performed:

1. The agent *DistributedQueryAgent* is created.
2. The *DistributedQueryAgent* verifies in the repository, how many ContainerAgents are present and gets their list. A request is sent to all considered instances to make their ontologies accessible for the integration process.
3. *ContainerAgent* in the moment of receiving a request from DistributedQueryAgent creates in its original instance the *DistributedIntegratingAgent* (E) and transfers a request to it.
4. *DistributedIntegratingAgent* picks the ontology stored in its own origin instance Then it sends it back to the *AgentContainer*.
5. *ContainerAgent* prepares a new message which contains the obtained ontology and sends it to a proper *DistributedQueryAgent*.
6. *DistributedQueryAgent* sends subsequent incoming ontologies to *QueueAgent* which places them into the queue.
7. The integration is executed. Only one integration at the same time may be performed, to guarantee the coherence of the knowledge. Ontologies present in the queue are subsequently sent to the created *GradeAgent* which, on the basis of the ontology stored in its own instance and obtained in the message, performs an estimation and performs the integration algorithms.
8. The last queue sends a confirmation by *GradeAgent* which signifies the end the integration and allows then to start the next integration process by taking the ontology from the queue managed by *QueueAgent*.

**3.3 Integration Process**

The process of integration (Fig. 2) consists of the following steps:

- Receiving of information containing the ontology models. After its initialisation an agent is in the state of waiting for information which orders it to start the process of estimation and integration. After receiving this information, an agent unpacks it and obtains the ontology models and a queue which should be performed by it.
- Initialisation of agents needed to perform estimations and integration. During the initialisation of GradeAgent it obtains a list of agents, which should be used. On the basis of this list, they are initialised and waiting for the orders of GradeAgent.
- Estimation procedure – a next step is to order the agents to estimate the similarities between concepts/classes. All possible combination of classes from both considered ontologies are checked with the use of the selected methods.
- Each integrating agent sends a matrix with the evaluations of similarities. The Grade Agent is looking for the best adjusted classes in the matrixes. As a results it obtains a list of integration commands describing the mode of integration of given concepts (copying or merging).

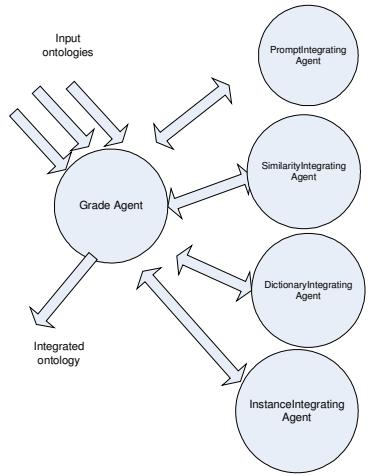

Fig. 2. Integration process

- On the basis of the list of commands a final integration process is performed. As a result, an integrated model is obtained.

## 4. Example of Application Domain: Supply Chains Management

Ontology-based approaches also seem to be a perfect solution for modelling and optimizing production processes. The amount of participants of such processes (customers, suppliers, producers etc) as well as a complexity and depth of relations between all participants of such processes, the number of parameters, coefficients and factors that have to be taken into consideration and finally all the above, causes supply chain management and optimization, production process management and optimization to be still a challenging task for contemporary algorithms, tools, representations and methods. The authors attempted to apply such ontology-based processes for modelling and optimization and below, one of the proposed and preliminary assessed approaches presented and discussed in short.

**4.1 Ontologies**

The developed system operates on the knowledge describing products, producers and orders. Logically, it can be separated into the part describing factories and products and into the part describing orders. With both of these parts there is a separated ontology associated. In order to realise the goal included in our research, the system is responsible for searching factories being able to produce commodities in a expected time, price, quality etc or for searching for substitutes of given product(s) if necessary and a factory being able to produce it.

As one of our goals was also to present the ability of integrating the domain-consistent data - the idea of dividing ontology into two separated ontologies is even more so justified.

The ontology describing factories and products includes the following information: the hierarchy of factories, categorisation of products; detailed hierarchy of products; localization of factories, information about products stored in database, qualitative and quantitative specifications of semi-products of a given product.

Next, the ontology describing orders includes the following information: the hierarchy of products and information about orders. In the proposed system, orders are visualized as an individual class of order. They are described by the orders group of parameters. With each order there is the number of ordered units associated. Next, ordered units are presented in the system as instances of appropriate class.

**4.2 Architecture**

From the architectural point of view there can be distinguished three main modules in the system (fig. 3.) : model description module; data module; operation performing module. Model description module enables and is responsible for adding information about individuals coming from data module (with the use of JENA library). It is able for operating on any ontology but in presented system and application it includes obviously mentioned ontologies describing factories, commodities and orders.

Data module includes information about individuals representing concepts from model description module. It is realised as a database implementation (with the use of MySQL server). The operation of mapping between ontology and database notions can be realised with the use of many different databases and it is realised thanks to D2RQ-based approach.

Operation performing module supplies input and output for any operation performed by the system. It is implemented in JAVA and it realizes reasoning on the data set coming from model description module with the use of mechanisms supplied by PELLET library. Operation performing module uses the interface of performing queries of SPARQL supplied by JENA library. It also supplies the mechanism of D2RQ by JENA API. Data coming from model description module are transformed to the JENA model and next with the use of PELLET library it is possible to have access to the processed and reasoned data.

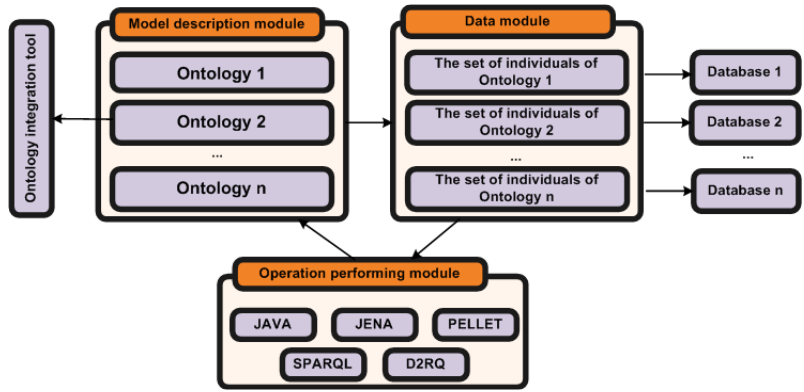

Figure 3. High-level system architecture

**4.4 Discussion of results**

One of the aspects of our research was to assess the ability, flexibility, simplicity and efficiency of knowledge integration. As mentioned before, ontologies used in our model were divided into two separate ontologies describing *factories/products* and *orders* respectively. What we wanted to do was to assess if their integration is possible and if so – how easy and how efficient or how difficult it is. During performed experiments dedicated tool developed in our group was used. The tool can operate in two modes: simple and full mode. In simple mode the user is able to define ontologies to integrate as well as some basic options – i.e. the user has to define if ontologies that should be integrated are dissimilar both lexically and structurally, if they are similar lexically but dissimilar structurally, if they are similar structurally and dissimilar lexically and finally if they are similar both: lexically and structurally. In full mode the user has to define among the others: the measure of the similarity (lexical, structural, global etc.), the algorithm and the filter of the similarity that should be used.

With the use of the above-mentioned tool we tried to integrate ontologies describing *factories and products* and *order*.

To estimate the quality of the performed integration measures of unconditional (defined as ratio of correct adjustments performed by the application to expected correct adjustments) and conditional quality success (defined as a ratio of correct adjustments performed by the application to all obtained adjustments). As a result of the considerable similarity of the subclasses of the Factories and Products Classes which have convergent names and similar internal structure, excessive adjustments are present.

## 5. Conclusions

In this work, an overview of methods of ontology integration and an agent infrastructure were presented, which uses these methods to integrate distributed knowledge sources. As an integration algorithm a similarity flooding was chosen since it allows for distinguishing points where classes that are being integrated seem to be similar. Almost all expected matchings were realized. It turned out also that using a similarity flooding algorithm was an appropriate choice since products are aggregated in both ontologiess inside one super-class and as a result, more appropriate matchings were realized

Then we presented a domain of the application of the ontological approach, emphasizing its advantages (on the level of reasoning and integration) which it offers during analysis and which is necessary in specific problems and for specific domain knowledge.